\documentclass[11pt]{article}
\usepackage{moriond,epsfig}
\usepackage[latin1]{inputenc}
\usepackage[OT1]{fontenc}
\usepackage{amssymb}
\usepackage{subfigure}

\def\Journal#1#2#3#4{{#1} {\bf #2}, #3 (#4)}

\def\be{\begin{equation}}
\def\ee{\end{equation}}
\def\bea{\begin{eqnarray}}
\def\eea{\end{eqnarray}}

\def\pl{\textit{Planck}}
\def\xmm{\textit{XMM-Newton}}

\newcommand{\commentaire}[1]{}
\begin{document}
\vspace*{4cm}
\title{AN SZ/X-RAY GALAXY CLUSTER MODEL AND\\THE X-RAY FOLLOW-UP OF THE PLANCK CLUSTERS}

\author{A. CHAMBALLU$^{1,\, 2}$, J. G. BARTLETT$^2$, J.-B. MELIN$^3$, M. ARNAUD$^4$}

\address{$^1$ Astrophysics group, Blackett Laboratory, Imperial College, Prince Consort road,\\ London NW7 2AZ, England\\
$^2$ Laboratoire AstroParticule \& Cosmologie (APC), Universit\'{e} Paris Diderot,\\ 10, rue Alice Domon et L\'{e}onie
Duquet, 75205 Paris Cedex 13, France (UMR 7164)\\
$^3$ CEA, DSM, SPP, 91191 Gif sur Yvette, France\\
$^4$ CEA, DSM, SAp, 91191 Gif sur Yvette, France}
\commentaire{$^{3/4}$ CEA, DSM, SPP/SAp, 91191 Gif sur Yvette, France}

\maketitle\abstracts{Sunyaev-Zel'dovich (SZ) cluster surveys will become an important cosmological tool over next few years, 
and it will be essential to relate these new surveys to cluster surveys in other wavebands. We present an empirical model of
cluster SZ and X--ray observables constructed to address this question and to motivate, dimension and guide 
X--ray follow--up  of SZ surveys.  As an example application of the model, we discuss potential \xmm\ follow-up of  \pl\ clusters.}

\section{Introduction}

Galaxy clusters are powerful cosmological probes:  observations of their internal structure provide information
on dark matter and can be used to estimate distances, while studies of their evolution gauge the 
influence of dark energy on structure formation 
-- as rare objects at the top of the mass hierarchy, 
their number density and its evolution are extremely sensitive to the underlying cosmology. 
For these reasons, the Dark Energy Task Force included cluster surveys among the four primary methods for constraining 
dark energy.  In this context, massive clusters are the most pertinent 
because their properties are little affected  by non--gravitational processes.

Today, we do not yet have the large samples of clusters, most notably massive clusters, out to high redshifts (e.g, unity and 
beyond) needed to fully realize the potential of cluster studies.  This is changing, thanks in large part to Sunyaev-Zel'dovich (SZ) 
cluster suveys.  The SZ effect\,\cite{sz1,sz2,sz3} is a distortion of the cosmic microwave background (CMB) black body 
spectrum due to inverse Compton scattering of CMB photons off electrons in the intra--cluster 
medium (ICM).  It is one of the most promising ways of finding new galaxy clusters, since its amplitude (in terms of surface
brightness) and spectrum are independent of redshift (in the non-relativistic case). 
\commentaire{The change in the intensity of the CMB induced by the SZ effect follows:
\be
\frac{\Delta I_{\nu}}{I_0} = y f(\nu) = \int_{los}\frac{kT_e}{m_ec^2}n_e\sigma_Tdl \times \frac{x^4e^x}{(e^x-1)^2}\left[\frac{x(e^x+1)}{e^x-1}-4\right]\, ,
\label{eq:delta_i_sz}
\ee
where $I_{\nu} = I_0\frac{x^3}{e^x-1}$ is Planck's law with $I_0 = \frac{2(kT_{\mbox{\tiny{CMB}}})^3}{(hc)^2}$, $T_e$ and $n_e$ are the temperature and density of the electrons in the ICM and  
$x = \frac{h\nu}{kT_{\mbox{\tiny{CMB}}}}$ in the dimensionless frequency ($\nu$ is the frequency of observation). The amplitude of the effect is characterized by the so-called \textit{Compton-y parameter}, which corresponds 
to the integral of the pressure along the line of sight (los). The spectrum of the distorsion, characterised by $f(\nu)$ in Eq. \ref{eq:delta_i_sz} is completely independent of the cluster's properties (at least, 
in the non relativistic case) and is thus universal; the distorsion corresponds to a deficit of photons relative to the mean sky ($\Delta I_{\nu}<0$) for $\nu \lesssim 217$~GHz, an excess 
($\Delta I_{\nu}>0$) for $\nu \gtrsim 217$~GHz and is null for $\nu\simeq 217$~GHz. 
Except for the changes of angular size on the sky, two identical clusters at, for instance, redshifts $0.5$ and $5$ will be responsible for the same distorsion.}
As SZ surveys begin to open this new window on cluster science, relating them to surveys in other wavebands becomes a 
critical issue, both to understand what we are finding and to fully exploit their scientific potential.

\section{An SZ/X--ray Cluster Model}

Observations of the SZ effect give only two--dimensional information projected onto the sky, and follow--up in other 
wavebands is essential for most studies.  Obviously, optical/NIR follow--up is needed to obtain redshifts.  
Much  can also be gained by combining X--ray and SZ data sets, for instance to better understand various 
survey selection functions.  Follow--up with \xmm\ and \textit{Chandra} will enable us to probe the ICM with 
unprecedented precision, e.g, its thermal structure and the gas mass fraction.  Moreover, with X--ray data we can
estimate cluster masses through application of hydrostatic equilibrium.  

In order to inform such SZ/X--ray comparisons and follow--up of SZ surveys, we have constructed an empirical and 
easily adaptable model\,\cite{wam} relating the SZ and X--ray properties of clusters. 
This section briefly summarizes its most relevant aspects.

\subsection{Description of the model}

Our model is based on several ingredients, derived from observations, numerical simulations and theory. 
We employ scaling laws in order to relate observed properties with the fundamental cluster parameters, mass 
and redshift: the $M_{500}-T$\,\cite{m500_1,m500_2}, $L_X-T$\,\cite{l-t} and  
$f_{gas}-T$\,\cite{fgas} relations. The evolution of all these scaling relations is still poorly constrained. 
However, recent observations\,\cite{evol_mt,evol_lt} 
indicate that self--similar evolution tends to reproduce well the data.  Given this, we adopted self--similar 
evolution in all cases, and we subsequently validated this choice (see next section).

We approximate the spatial structure of the gas with an isothermal $\beta$-model\,\footnote{This will be 
improved in a future version of the model, to account for recent observations showing that this profile is inadequate, 
especially in the core and outer parts of clusters.} with $\beta=2/3$.  Fitting the $L_X-T$ relation then 
requires a deviation from self--similarity in the $f_{gas}-T$ relation, i.e., the gas mass fraction varies
with cluster total mass; this variation is allowed by present observations.  
For the dark matter, we adopt a NFW profile and use the Jenkins mass function\,\cite{mf}.  Local cluster counts in terms 
of the X-ray Temperature Function\,\cite{xtf} (XTF) then fix the normalization of the fluctuation power spectrum
(using the measured $M_{500}-T$).   

By combining these different ingredients we constrain all free parameters of the model, namely 
those describing cluster physics -- like the core radius $r_c$ and the central electronic density $n_e$  --
and those describing population statistics, such as $\sigma_8$ (see below). 
We took particular care with the various mass definitions available in the literature, 
related to theoretical studies (e.g. $M_{vir}$), observations (e.g. $M_{500}$) or numerical simulations
(e.g. masses estimated by the friends--of--friends method), transforming among them with the 
NFW dark matter profile.  This was indispensable for coherently combining the variety of constraints. 

\subsection{Model validation}

\begin{figure}[t!]
\begin{center}
\subfigure{\epsfig{figure=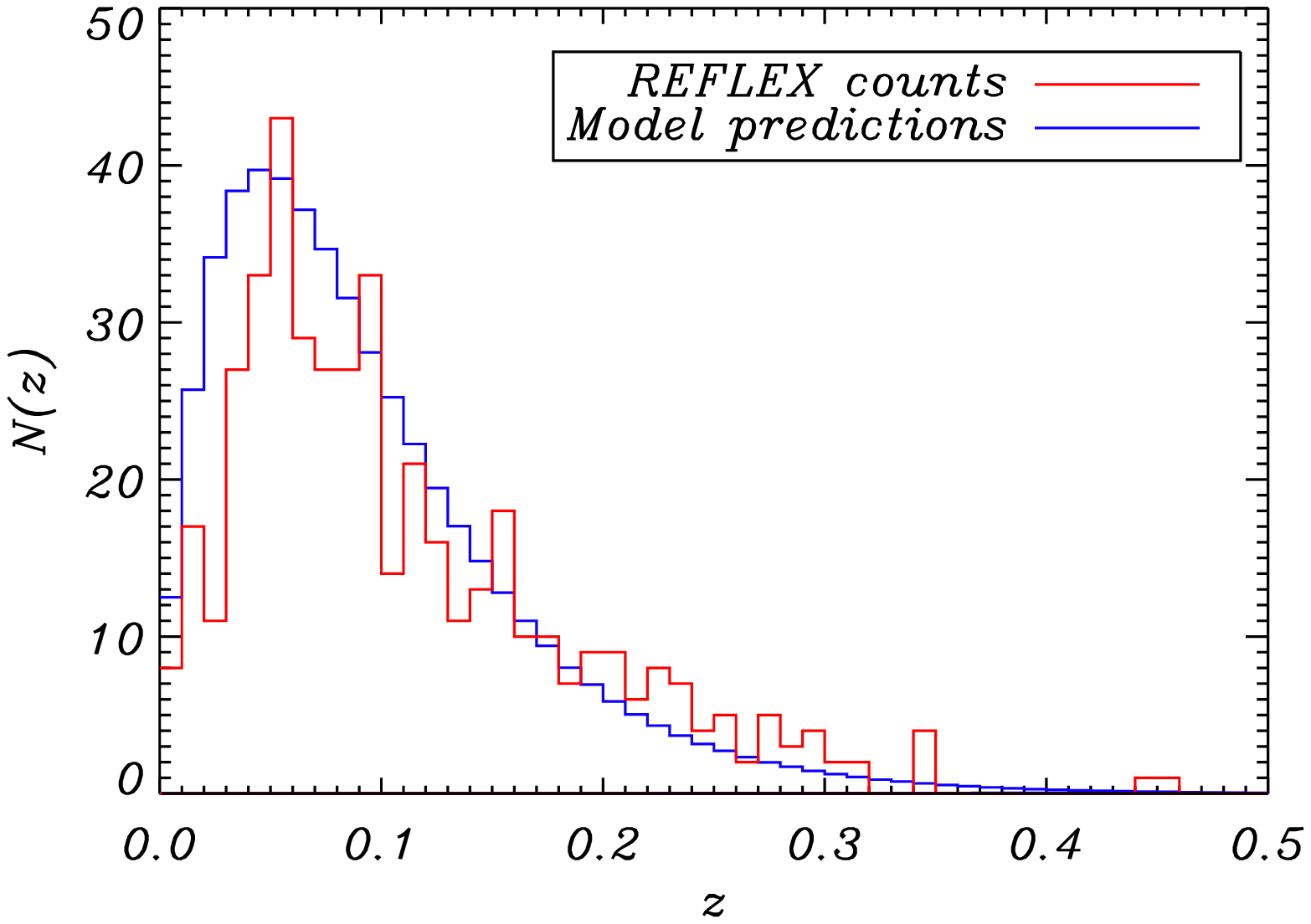,width=7.0cm}}
\subfigure{\epsfig{figure=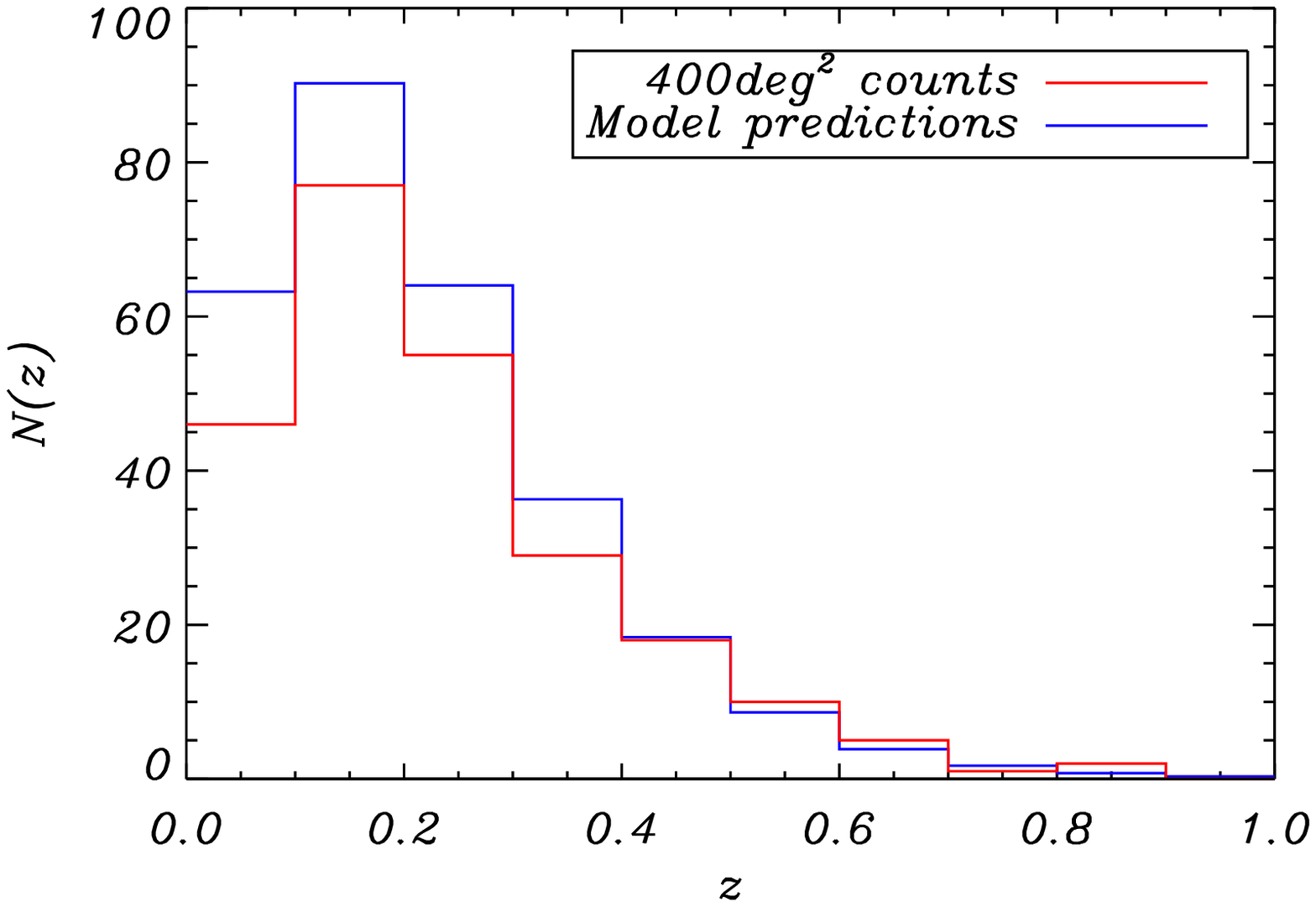,width=7.0cm}}
\end{center}
\vspace{-0.5cm}
\caption[The Planck catalog]{Two examples of the redshift distribution of clusters from ROSAT surveys (red) compared to model predictions (blue). 
{\em Left}: The REFLEX survey. {\em Right}: The 400 square--degree survey.}
\label{fig:xray_counts}
\end{figure}

To validate the model, we checked it against additional observational constraints, not used to 
fix its parameters.  We discuss below the most relevant of these\commentaire{, redshift distributions of X--ray
surveys}, but cite another notable one in passing: 
fitting 
the local XTF\,\cite{xtf} we find  $\sigma_8=0.78\pm 0.027$, in complete agreement with WMAP-5 
results\,\cite{wmap}.
More specifically, we tested the model by comparing the observed redshift distributions from the REFLEX\,\cite{reflex} and 400 square--degree\,\cite{400} surveys to the predictions of the model. 
Figure \ref{fig:xray_counts} shows the 
predicted and observed counts in both cases. In the former case, the observed total number of clusters is 447 
with a completeness estimated to be at least 90\%; the predicted number is 508 clusters, which corresponds 
to 457 clusters for a completeness of 90\%. Moreover, the shapes of the two distributions are in very good agreement. 

In the case of the 
400deg$^2$ survey, the model reproduces extremely well the high redshift distribution ($z>0.4$), although its seems to 
predict too many low redshift clusters. Noting that this is a serendipitous survey, in which known local clusters are by
construction missing, we conclude once again that the model is in reasonable agreement with the data. 
This last result is particularly satisfying since the high redshift clusters contained in this deep survey are of the kind 
expected to be found in SZ surveys like \pl\, (as discussed below).

\section{An application of the model}

The model is completely general and can be used to predict the results of any set of SZ and X--ray observations.  
As an example of its application, we discuss potential follow--up of  \pl\ SZ  clusters with \xmm. 

\subsection{The \pl\ cluster catalog}

To accurately model the  \pl\ cluster catalog, we employed the selection function derived using the detection 
algorithms developed by Melin et al.\,\cite{jb} and applied to detailed simulations of \pl\ observations (the \pl\
Sky Model\,\cite{psm}).  We find that a non--negligible fraction of otherwise bright SZ clusters remain undetected: these are resolved, 
low to intermediate redshift clusters whose SZ flux is diluted over several pixels. 
This selection effect is shown in the left--hand panel of Figure 
\ref{fig:catalog}, where we plot the cumulative redshift distribution of \pl\ clusters.  

The result is that the \pl\ catalog is expected to contain $\sim$2350 clusters\commentaire{ (blue curve)}, of which $\sim$180 
are at $z>0.6$ and $\sim$15 at $z>1$; there is, of course,  a certain amount of model uncertainty associated with these 
predictions, in particular from the normalization of the SZ--mass relation.  
%We now use our model to derive X--ray properties 
%of the Planck catalog and to dimension an X--ray follow-up program with \xmm. 

\subsection{Follow--up with \xmm}

\begin{figure}[t!]
\begin{center}
\hspace{-0.8cm}
 \subfigure{\epsfig{figure=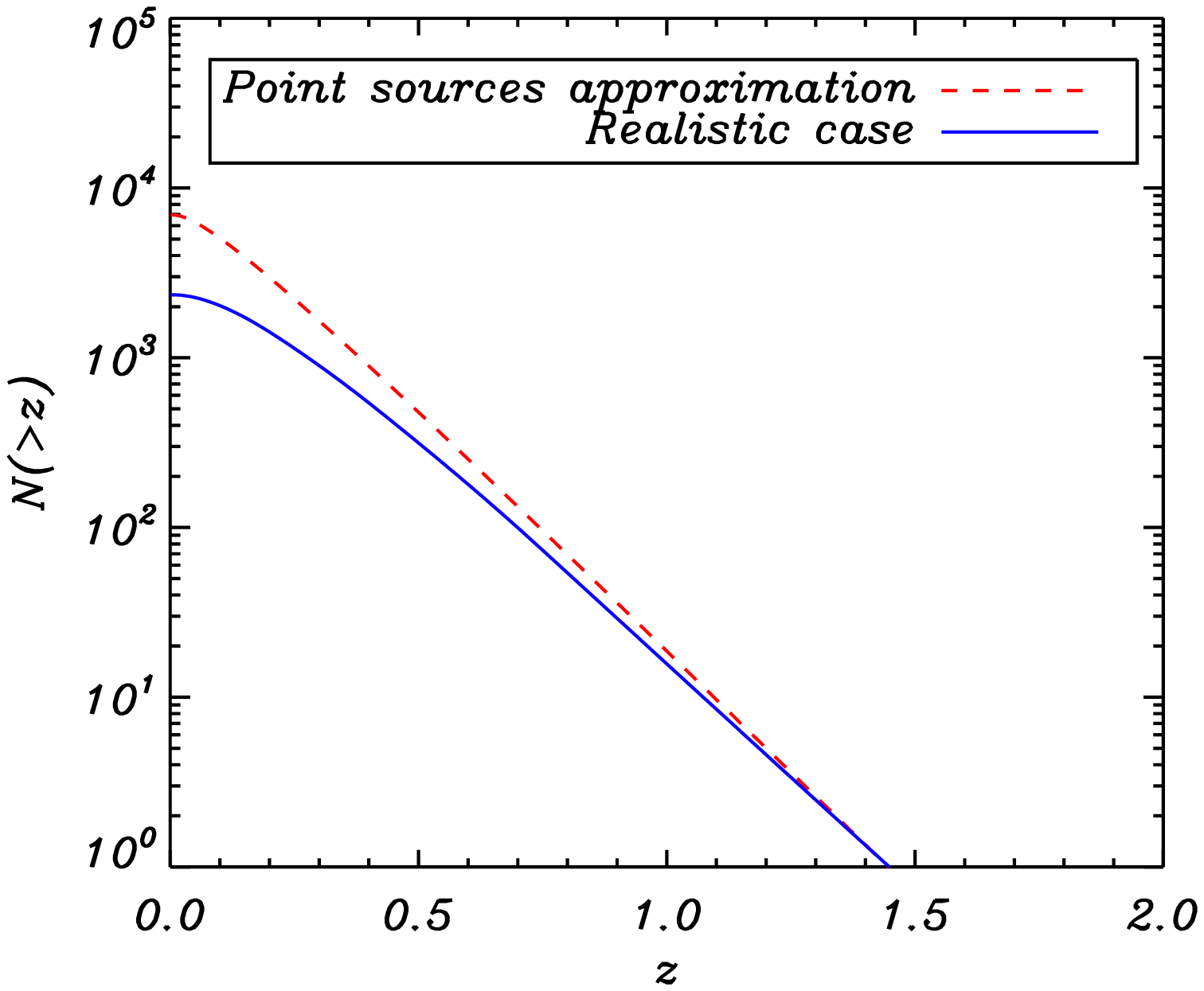,width=5.9cm}}\ %\\ \vspace{-1cm}
 \subfigure{\epsfig{figure=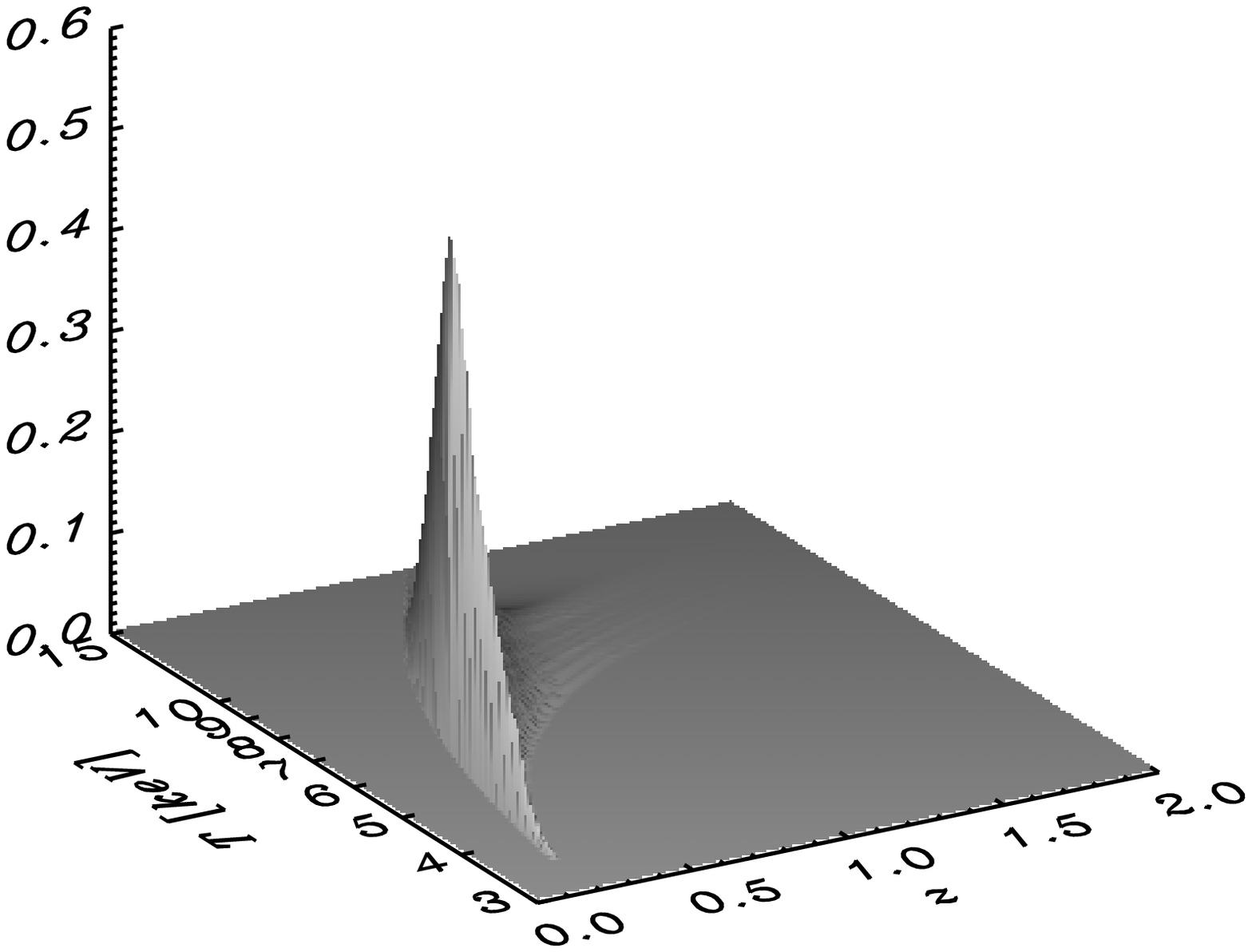,width=4.8cm, height=4.8cm}}\hspace{-0.5cm}%\quad
 \subfigure{\epsfig{figure=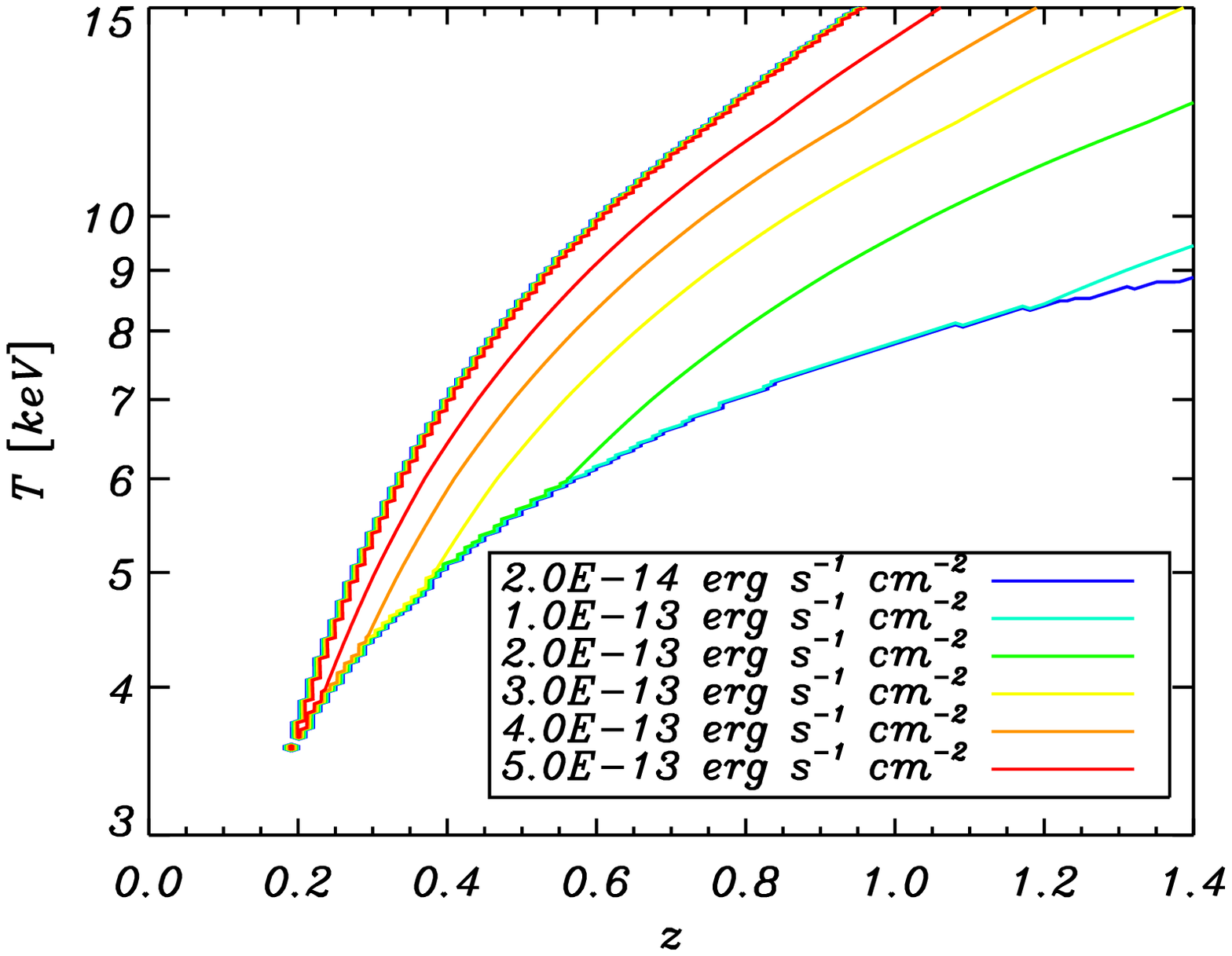,width=6.3cm}}
\end{center}
\vspace{-0.3cm}
\caption[The Planck catalog]{{\em Left}: Predicted cumulative redshift distribution of the {\it Planck} cluster catalog. The dashed red line corresponds to the case where all clusters are (falsely) imagined to be unresolved (point--source approximation); 
the blue line is the realistic case, accounting for the fact that some clusters are resolved. 
{\em Middle}: The {\it Planck} newly--discovered cluster catalog (in which clusters are 
observed by {\it Planck} but not by ROSAT) distributed in bins over temperature and redshift. {\em Right}: The same 
distribution projected over the ($z$, $T$)--plane with contours of iso--flux in the 
{\it XMM-Newton} [0.5-2]--keV band.}
\label{fig:catalog}
\end{figure}

We wish to identify the X--ray nature of these new \pl\ clusters and evaluate the ability of \xmm\ to observe a 
significant number of them.  We therefore examine those clusters with X--ray fluxes below the ROSAT All Sky Survey 
(RASS) limit, which we take to be  
$f_{X}[0.1-2.4]\mbox{keV}= 10^{-12}$ erg s$^{-1}$ cm$^{-2}$ (i.e., the lowest limit of the MACS survey\,\cite{macs}). 
The distribution of this sub--catalog of new \pl\ clusters is given as a function of redshift 
and predicted temperature in the central panel of Figure \ref{fig:catalog}. Most of the ($\sim$520) clusters are relatively
cool and local; however, $\sim$168 clusters lie at $z>0.6$ and have temperatures $T>6$keV.  Note that only six such
clusters are presently known.  

In the right--hand panel of  Figure \ref{fig:catalog}, we show the expected X--ray flux of these objects 
in the \xmm\  [0.5-2]--keV band as contours projected onto the redshift--temperature plane.
This allows us to evaluate their detectability, and we see that all of these \pl\ clusters have fluxes larger than 
$10^{-13}$ erg s$^{-1}$ cm$^{-2}$.  They are bright, falling in the flux decade just below the ROSAT limit.
This has important consequences for follow--up programs. 

Using observations of MS1054-0321\,\cite{ms} and 
ClJ1226.9+3332\,\cite{clj} -- two clusters of the same kind as the newly--discovered high redshift \pl\ clusters -- as a guide, we 
estimate that \xmm\ could measure the temperature of \pl\ clusters at $z>0.6$ to 10\% with a relatively short
exposure of 25-50 ks (per cluster).  It should also be possible to  obtain masses and mass profiles for the reasonably relaxed
 clusters by applying hydrodynamic equilibrium equation.  
 
\section{Summary}

We presented a model for the SZ and X--ray signals of galaxy clusters based on current X--ray data. 
Using a realistic mock  \pl\ cluster catalog, we employed the model to 
predict, firstly,  that $\sim$168 newly--discovered clusters lie at $z>0.6$ with \commentaire{temperatures }$T>6$keV, and secondly,
that these clusters can be observed in some detail in only 25--50 ks with \xmm.   
Thus
%This means that 
we could follow--up the majority of these new \pl\ clusters with a dedicated program of several Msec on \xmm; 
this falls in the category of {\em Very Large Programme} now possible with the satellite.  Follow--up
observations with \xmm\ would therefore dramatically increase the sample of well--studied, 
massive, high redshift ($0.6<z<1$) clusters, key objects for precise cosmology with clusters
and for testing gravitational processes in cluster formation.
\commentaire{and thereby move the kind of studies performed in the local universe well out in redshift.}

\end{document}